\documentclass[twocolumn]{aastex63}
\usepackage{graphicx}
\usepackage{amsmath}
\usepackage{amssymb}
\usepackage{cases}
\usepackage{mathrsfs}
\usepackage[flushleft]{threeparttable}
\usepackage{amssymb}
\usepackage{pifont}
\usepackage{amsmath}
\usepackage{epstopdf}
\usepackage{xcolor}
\usepackage[all]{hypcap}
\usepackage{mwe,tikz}
\usepackage{bm}
\usepackage{tabularx}
\usepackage{longtable}
\usepackage{booktabs}
\usepackage{xtab}
\usepackage{CJK}
\usepackage{xspace}
\usepackage{rotating}

\newcommand{\e}{\text{e}}

\newcommand{\p}{\partial}



\shorttitle{Search for X-ray Bursts in SGR J1935+2154}
\shortauthors{Zou et al.}

\def \scalgleP {$\sim$238}
\def \zjhP {238}

\begin{document}

\title{Periodicity Search on X-Ray Bursts of SGR J1935+2154 Using 8.5 yr of Fermi/GBM Data}
\correspondingauthor{Bin-Bin Zhang; Lang Shao}
\email{bbzhang@nju.edu.cn,lshao@hebtu.edu.cn}

\author[0000-0003-4011-2608]{Jin-Hang Zou}
\affiliation{College of Physics, Hebei Normal University, Shijiazhuang 050024, China}
\affiliation{School of Astronomy and Space Science, Nanjing
University, Nanjing 210093, China}

\author[0000-0003-4111-5958]{Bin-Bin Zhang}
\affiliation{School of Astronomy and Space Science, Nanjing
University, Nanjing 210093, China}
\affiliation{Key Laboratory of Modern Astronomy and Astrophysics (Nanjing University), Ministry of Education, China}
\affiliation{Department of Physics and Astronomy, University of Nevada Las Vegas, NV 89154, USA}

\author[0000-0001-6545-4802]{Guo-Qiang Zhang}
\affiliation{School of Astronomy and Space Science, Nanjing
University, Nanjing 210093, China}
\affiliation{Key Laboratory of Modern Astronomy and Astrophysics (Nanjing University), Ministry of Education, China}

\author[0000-0003-0691-6688]{Yu-Han Yang}
\affiliation{School of Astronomy and Space Science, Nanjing
University, Nanjing 210093, China}
\affiliation{Key Laboratory of Modern Astronomy and Astrophysics (Nanjing University), Ministry of Education, China}

\author[0000-0001-8876-2357]{Lang Shao}
\affiliation{College of Physics, Hebei Normal University, Shijiazhuang 050024, China}

\author[0000-0003-4157-7714]{Fa-Yin Wang}
\affiliation{School of Astronomy and Space Science, Nanjing
University, Nanjing 210093, China}
\affiliation{Key Laboratory of Modern Astronomy and Astrophysics (Nanjing University), Ministry of Education, China}

\begin{abstract}

We performed a systematic search for X-ray bursts of the SGR J1935+2154 using the Fermi Gamma-ray Burst Monitor continuous data dated from 2013 January to 2021 October. Eight bursting phases, which consist of a total of {353} individual bursts, are identified.
{We further analyze the periodic properties of our sample using the Lomb-Scargle periodogram. The result suggests that those bursts exhibit a period of $\sim$ \zjhP\ days with a $\sim$ 63.2\% duty cycle}. Based on our analysis, we further predict two upcoming active windows of the X-ray bursts. Since 2021 July, the beginning date of our first prediction has been confirmed by the ongoing X-ray activities of the SGR J1935+2154.

\end{abstract}

\keywords{Gamma-ray bursts; Soft gamma-ray repeaters; Magnetars; Gamma-ray transient sources}

\section{Introduction} \label{sec:intro}

Soft gamma-ray repeaters (SGRs), the X-ray/gamma-ray transient sources that exhibit explosive activity about every year or decade, are known to be magnetars \citep{1995ApJ...444L..33V,1997ApJ...480..607B,1999A&A...350..891F,2000ApJ...533L..17V}, the young neutron stars with strong magnetic fields \citep{1992ApJ...392L...9D,1992AcA....42..145P,1995MNRAS.275..255T,1996ApJ...473..322T,1998Natur.393..235K}. Such a magnetar origin of the SGRs has been confirmed by a few recent cases such as SGR 1806-20 \citep{1981Ap&SS..75...47M}, SGR 0526-66 \citep{1979SvAL....5...87M,1979Natur.282..587M}, SGR 1627-14 \citep{1998IAUC.6944....2K,1999ApJ...519L.139W}, SGR 1900+14 \citep{1979SvAL....5..343M,1993Natur.362..728K}, and SGR J1935+2154 \citep{2014GCN.16520....1S}, all of which are listed in the magnetar catalog\footnote{\url{http://www.physics.mcgill.ca/~pulsar/magnetar/main.html}} \citep{2014ApJS..212....6O} without any counterexample.

According to their brightness, the SGR bursts can be roughly divided into three classes \citep{2006csxs.book..547W}:
\begin{itemize}
 \item Short-duration bursts: the most common type of bursts with typical duration of about $~0.1\,{\rm s}$ and spectra characterized by the Optically Thin Thermal Bremsstrahlung (OTTB) model. 
 
\item Giant flares (GFs): unusual intense bursts with energy about a thousand times higher than that of a typical burst, characterized by an initial hard initial spike and the rapidly decaying tails with pulsations \citep[e.g.][]{1979Natur.282..587M,1999Natur.397...41H}. 
Some short gamma-ray bursts (GRBs), namely, GRB 051103 \citep{2006ApJ...652..507O,2007AstL...33...19F}, GRB 070201 \citep{2008ApJ...680..545M, 2008ApJ...681.1464O}, GRB 200415A \citep{2020ApJ...899..106Y, 2021Natur.589..211S}, and a few more, are indeed found to be magnetar-giant-flare (MGF) originated \citep{2021ApJ...907L..28B}.

\item Intermediate bursts (IBs): intermediate bursts with peak flux, duration, and energy between the short-duration bursts and GFs \citep[e.g.][]{1984Natur.307...41G,2004A&A...416..297G}. They tend to have abrupt onsets as well as abrupt endpoints \citep{2006csxs.book..547W} if the duration is less than the rotation period ($\sim$ a few seconds) of the magnetar. 

 \end{itemize}
 
Multiwavelength afterglows are observed in both IBs and GFs. For example, a radio afterglow event \citep{2005Natur.434.1112C} has been observed from SGR 1806-20 after its GF in 2005 January \citep{2005GCN..2936....1B}. Two X-ray afterglow events have been observed from SGR 1900+14, following by its GF in 1998 August and IB in 2001 April respectively \citep{2001ApJ...552..748W,2003ApJ...596..470F}, and a radio afterglow has been observed following the GF of the same SGR in 1998 August \citep{1999Natur.398..127F}.

SGR J1935+2154 is a Galactic magnetar, which was first observed by the Swift Burst Alert Telescope (BAT) in 2014 \citep{2014GCN.16520....1S}. It has experienced four active windows before 2020, respectively, in 2014, 2015, 2016, and 2019 \citep{Younes_2017,2020ApJ...902L..43L}. 2020 April was recognized as the most violent bursting month of SGR J1935+2154 so far, during which a burst forest was observed. These bursts consist of the first X-ray counterpart \citep{2021NatAs.tmp...54L,2020ATel13686....1T,2020ApJ...898L..29M,2021NatAs...5..372R} that is associated with a fast radio burst, FRB 200428 \citep{2020Natur.587...59B,2020Natur.587...54C}.

FRBs are millisecond radio transients with large dispersion measures \citep[DMs $\sim 100-2600$ pc cm $^{-3}$;][]{Lorimer2007Sci...318..777L,Cordes2019ARA&A..57..417C,Petroff2019A&ARv..27....4P,Zhang2020Natur.587...45Z,Xiao2021}. 
Although hundreds of FRBs have been observed so far\footnote{\url{http://frbcat.org/}}, the physical models are still under debate \citep[for a recent review, see][]{Platts2019PhR...821....1P}. Multiple models suggest that FRBs can originate from a magnetar. Those models involve giant flares of a magnetar, the interactions between magnetar flares and their surroundings \citep{Kulkarni2014ApJ...797...70K,Lyubarsky2014MNRAS.442L...9L,Katz2016ApJ...826..226K,Beloborodov2017ApJ...843L..26B,Metzger2017ApJ...841...14M}, and the collision of a magnetar and an asteroid \citep{Geng2015ApJ...809...24G,Dai2016ApJ...829...27D}. Interestingly, the statistical properties of FRBs are similar as those of Galactic magnetar bursts \citep{Wang2017,Wadiasingh2019,Cheng2020}. The association between an X-ray burst of SGR J1935+2154 and FRB 200428 supports that at least some of FRBs are produced by magnetars \citep{2020Natur.587...54C,2020Natur.587...59B}. 

The magnetar association of FRBs motivated us to search for the common properties shared by the two phenomena. An interesting manner of repeating FRBs is the periodic window behavior (PWB), proposed in the bursts of FRB 121102 and FRB 180916.
The PWB describes a quasi-periodic phenomenon that bursting phases always appear periodically, but there is no periodicity for specific bursts. 
\citet{Chime/FrbCollaboration2020Natur.582..351C} first reported a possible period about 16.35 days for FRB 180916. They collected 38 bursts that occurred from 2018 September to 2020 February and located these bursts in a 5 day phase window.
\citet{Rajwade2020MNRAS.495.3551R} suggested a possible PWB of about 160 days for FRB 121102. This result was also confirmed by \citet{Cruces2021MNRAS.500..448C}. Many models have been proposed to explain those periods, including those involved with the precession of magnetars \citep{Levin2020ApJ...895L..30L, Zanazzi2020ApJ...892L..15Z},
the orbit motion of binary stars \citep{Ioka2020ApJ...893L..26I, Lyutikov2020ApJ...893L..39L}, the ultra-long period magnetars \citep{Beniamini2020MNRAS.496.3390B}, the luminous X-ray binaries \citep{2021ApJ...917...13S} and the magnetar-asteroid model \citep{Dai2020ApJ...895L...1D}.

PWBs have also been investigated for SGRs. \citet{Zhang2021ApJ...909...83Z} analyzed the bursts of SGR 1806-20, and estimated a possible period of about 395 days. \citet{2021PASP..133g4202G} suggested a possible PWB with a period of about 231 days for SGR J1935+2154 using the observations from IPN\footnote{\url{http://www.ssl.berkeley.edu/ipn3/masterli.html}} (Interplanetary Network)
instruments. The PWB was further studied by \citet{2021PhRvD.104b3007D} using the likelihood analysis.

In this Letter, We performed a systematic search for X-ray bursts of the SGR J1935+2154 using the Fermi Gamma-ray Burst Monitor \citep[GBM; ][]{2009ApJ...702..791M} continuous data dated from 2013 January to 2021 July, aiming to identify its PWB using an unbiased sample. In section \ref{sec:pulse_search}, we describe the search procedure and report the results of our search. In section \ref{sec:physical_origin}, we use the Lomb-Scargle method to analyze the results and show that SGR J1935+2154 has a bursting period of about 238 days. Finally, brief implications and discussions are provided in Section \ref{sec:summary}.

\section{Burst Search} \label{sec:pulse_search}

Due to the limit of detector sensitivity, not all SGR bursts can trigger Fermi/GBM. Therefore, an untriggered search is needed to identify all potential bursts throughout the available data \citep{2020ApJ...902L..43L}.
Many searches have been carried out \citep[e.g.][]{2020ApJ...893..156L,2020Natur.587...63L,2020ApJ...904L..21Y,2020ApJ...898L..29M,2021ApJ...906L..12Y}. We independently searched the continuous time-tagged event (CTTE) data of Fermi/GBM NaI detectors within the range of 8 - 900 keV from 2013-01-01T00:00:00 (UTC) to {2021-10-21T23:59:59} (UTC), including the latest explosive activity of SGR J1935+2154 in {July} by the following three steps:

 \begin{figure*}
\label{fig:eg_lc}
\centering
\includegraphics[width=0.75\textwidth]{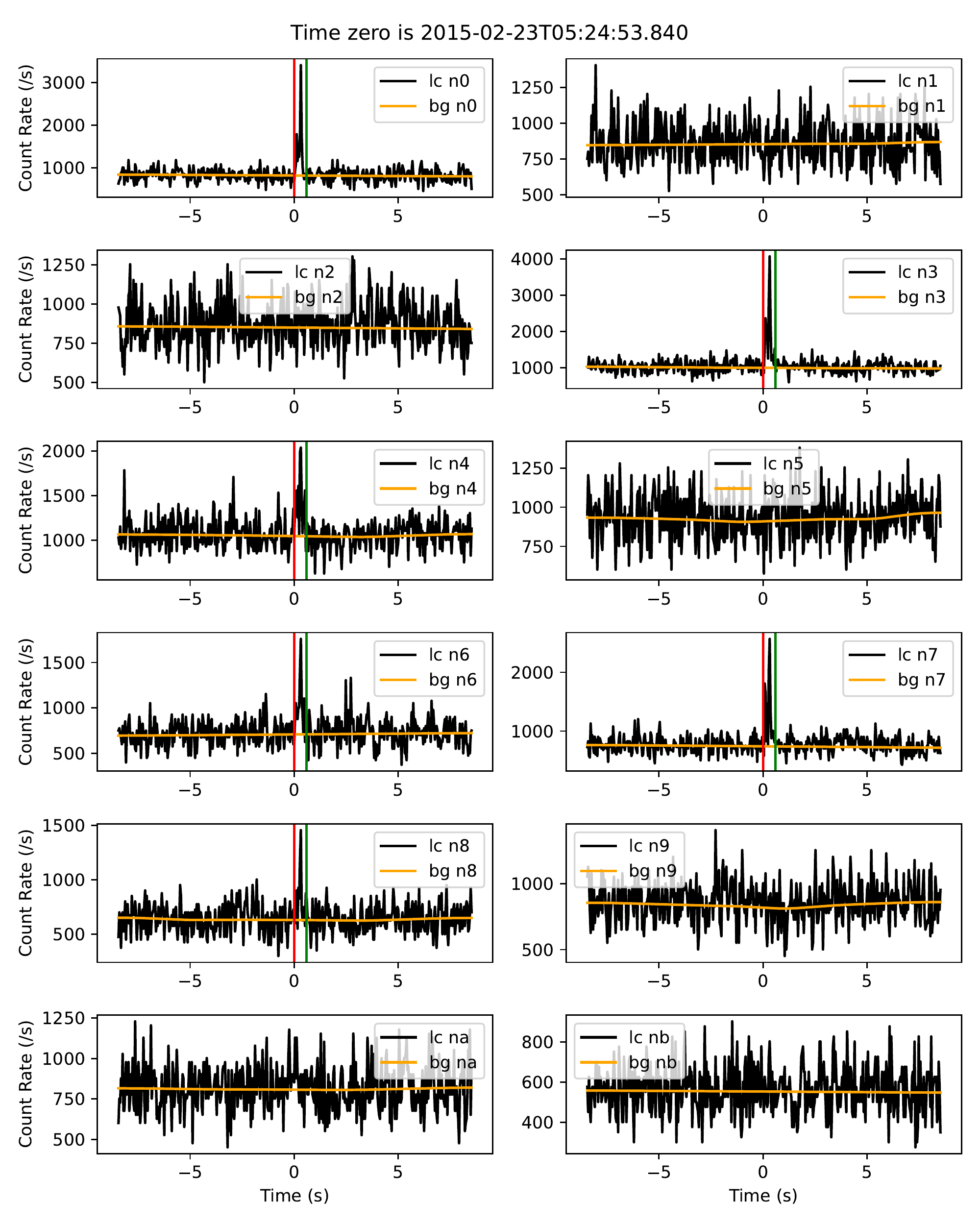}
\caption{ An example of a short-duration burst in our sample. The black curves represent the light curves (lc) of 12 NaI detectors, and the orange lines represent the background (bg) of each light curve. The red lines and the green lines represent the starting time and the ending time of the burst recorded in each detector, respectively. The starting time of this burst is 2015-02-23T05:24:53.840 (UTC).}
\end{figure*}

\begin{enumerate}

\begin{figure*}
\centering
\includegraphics[width=0.7\textwidth]{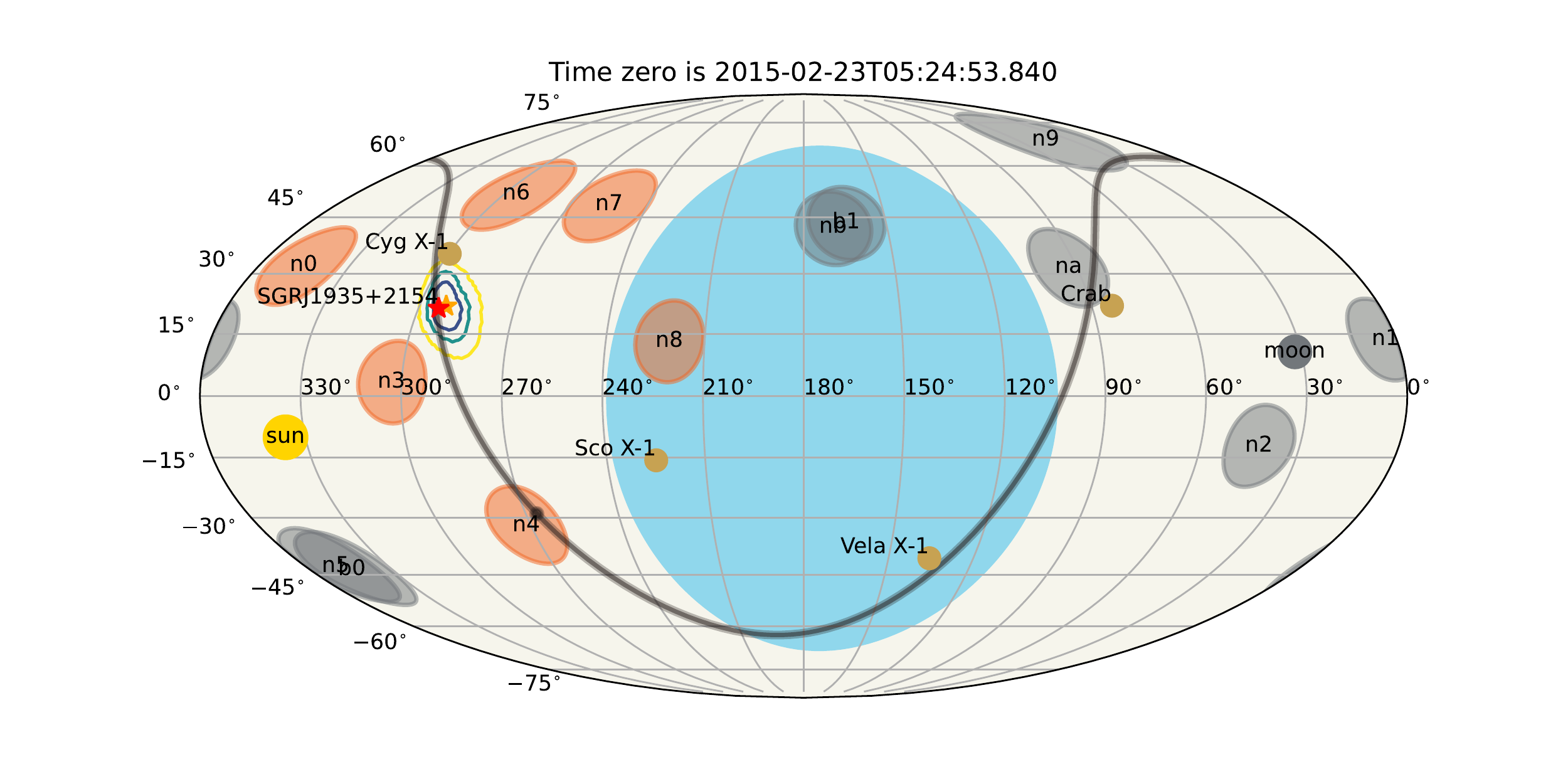}
\caption{The localization of the same burst in Figure \ref{fig:eg_lc}. The small circles represent the pointing directions of NaI detectors, {which can be calculated from the from satellite attitude data available in the Fermi/GBM data products}. The detectors illuminated and nonilluminated by SGR J1935+2154 are filled with orange and gray colors, respectively. The red star indicates SGR J1935+2154, and the yellow star indicates the location calculated by our code. The three contour lines in different colors represents the 1$\sigma$, 2$\sigma$, and 3$\sigma$ credible regions, respectively. The brown dots mark the locations of four bright X-ray sources, namely Cygnus X-1, Sco X-1, Vela X-1, and Crab. The black curve represent the projection of the galactic plane. The blue area represents the sky region occulted by the Earth.}
\label{fig:eg_skymap}
\end{figure*}

\begin{figure}
 \label{fig:A_map}
 \centering
 \includegraphics[width=0.49\textwidth]{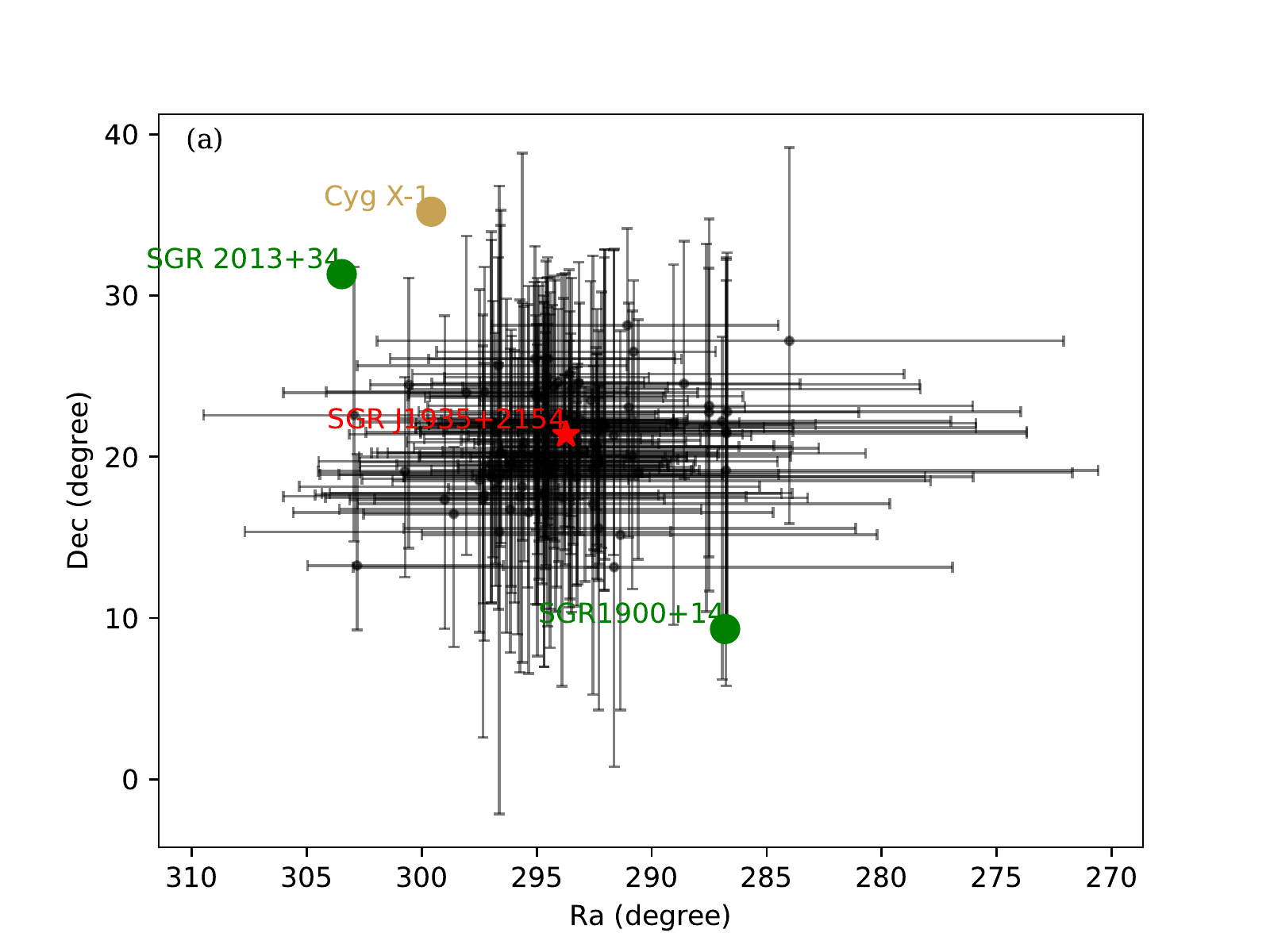}
 \includegraphics[width=0.49\textwidth]{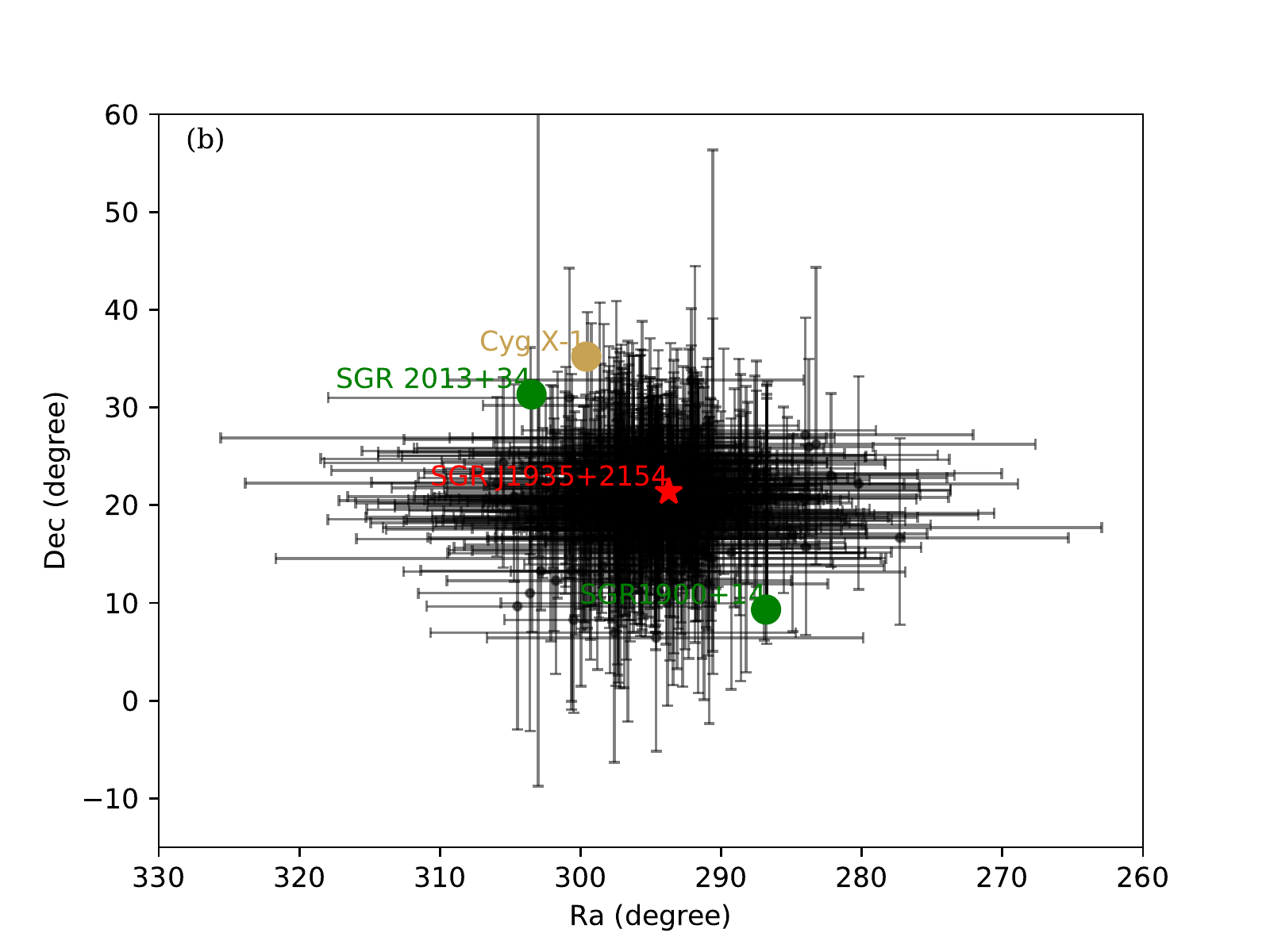}
 \caption{The top panel (a) shows the localization of 121 confirmed bursts of SGR J1935+2154 observed by Fermi/GBM. The red star represents the coordinates of SGR J1935 + 2154. The two green points are SGR 2013+34 and SGR 1900+14. The brown point is Cygnus X-1. Black dots and lines represent our localization and 1$\sigma$ localization error. The bottom panel (b) shows the localization of all the bursts in our sample.}
\end{figure}

 \item \textit{Data reduction and burst window selection.}
 
 The first step is to determine the data search windows in CTTE data that exclude the South Atlantic Anomaly (SAA) and contains possible bursts. The SAA passage can be easily determined by calculating the Fermi's longitude and latitude through the position history files released on Fermi/GBM data\footnote{\url{https://heasarc.gsfc.nasa.gov/FTP/fermi/data/gbm/daily/}}. To further determine the searching window, we first downloaded the hourly separated CTTE data of 12 sodium iodide (NaI) scintillators on board Fermi/GBM, binned them into 40-ms bin size, and obtained the corresponding light curves\footnote{We did not utilize the data from the two cylindrical bismuth germanate (BGO) scintillators on board Fermi/GBM, mainly because the X-ray bursts in our study are presented in the energy range that is mostly detectable by NaI detectors.}. The time ranges with zero count rate (e.g., due to the SAA passage) are excluded. The hourly data may be cut into one more pieces due to the SAA passage or missing data. Then, for each continuous piece, we constructed a background model using the baseline method implemented in R language\footnote{\url{https://www.rdocumentation.org/packages/baseline/versions/1.3-1}}. 
 Based on the modeled background, we can further select the qualified background range using sigma clipping method\footnote{\url{https://docs.astropy.org/en/stable/api/astropy.stats.sigma\_clip.html}}, within which the observed count rates are consistent with the background level. Based on the Gaussian fit of the count rate within the background range, we can further determine the 3$\sigma$ confidence level of the background. Finally, the burst search windows can be determined by bracketing the regions where $C_{obs,i} - B_{i}\ge 3\sigma$, which are used in step 2 below.

\item \textit{Burst identification by Bayesian block method.} 

In this step, we apply the Bayesian blocks method \citep{2013ApJ...764..167S}
to the burst windows found in Step 1 and identify burst candidates based on their temporal properties. 
The Bayesian blocks (BB) method\footnote{\url{https://docs.astropy.org/en/stable/api/astropy.stats.bayesian\_blocks.html\#astropy.stats.bayesian_blocks}} is a nonparametric modeling technique for detecting and characterizing local variability in time-series data \citep{2013ApJ...764..167S}. It maximizes the likelihood to find the best segmentation or boundary between blocks, called the ``change point", and has been used in untriggered searches in some previous studies \citep[e.g.][]{2020ApJ...902L..43L,2020ApJ...893..156L,2020Natur.587...63L,2021ApJ...906L..12Y}. 

For each burst window, we exact the light curves of 12 NaI detectors. Each light curve is ensured to cover at least three times the window width. We used the change points obtained by the BB method to divide it into ``blocks". Each of those blocks, by definition, can be characterized as having a constant block rate (``block rate"), which can be calculated by averaging the rate within the block. The longest block, which is also automatically the lowest one, provides us the first ``background block" to measure the background level. Ordered by their lengths, two to four additional blocks are checked and selected as background blocks if their block rates are consistent at the 1$\sigma$ level with that of the longest one. The final background region contains 2-5 background blocks. 

Next, an array of Boolean values is assigned with ``True'' for the light-curve data within the background region, and ``False'' otherwise. The background level of the whole burst window can then be calculated using the Whittaker Smoother method \citep{whittaker_1922,2003AnalChem}.

Finally, the background level can be used to calculate the significance, $S$, of each block according to Equation~(15) in \cite{2018ApJS..236...17V}. If $S$ exceeds the threshold, $3$, the block is recorded as a burst block candidate. We require that a qualified burst block must present in at least two NaI detectors. Eventually, all the burst blocks form a ``burst" candidate, and the spreading length of the burst blocks defines the burst duration. By requiring $S > 3 $, more than 1000 burst candidates are generated in this step. An example of our burst candidates is shown in Figure \ref{fig:eg_lc}.

\item \textit{Burst Confirmation by Localization.} 

The burst candidates are then localized and checked if they consistently conform with the location of SGR J1935+2154 at the 1$\sigma$ level.
We developed our own localization code following the method in \cite{Connaughton_2015} and \cite{2018MNRAS.476.1427B}. By employing the response matrix generator \textit{gbm\_drm\_gem}\footnote{\url{https://github.com/grburgess/gbm_drm_gen}} and the related database, our code can calculate the expected modeled count rate ratios among 12 NaI detectors for any directions. The burst location can be determined by maximizing the likelihood ratio between the modeled and observed data. An example of our localization results is shown in Figure \ref{fig:eg_skymap}. To verify the validation of our localization method, we selected the recent 121 confirmed bursts in previous studies \citep{2020ApJ...902L..43L,2020Natur.587...63L,2020ApJ...904L..21Y} and plot our location results for them in Figure \ref{fig:A_map} (a). We found that the 1$\sigma$ distance between the SGR and our locations is an adequate one to claim the association between our bursts and the SGR 1935+2154. The averaged 1$\sigma$ uncertainties of this sub-sample is 8.39 degrees. In addition, we plotted the locations of all the bursts in our sample in Figure \ref{fig:A_map} (b) and found the 1$\sigma$ measurement could also apply to other bursts. In particular, the averaged 1$\sigma$ value of the whole sample is 9.15 deg, which is consistent with the previous subsample.

\begin{figure*}
 \label{fig:summary}
 \centering
 \includegraphics[width=0.9\textwidth]{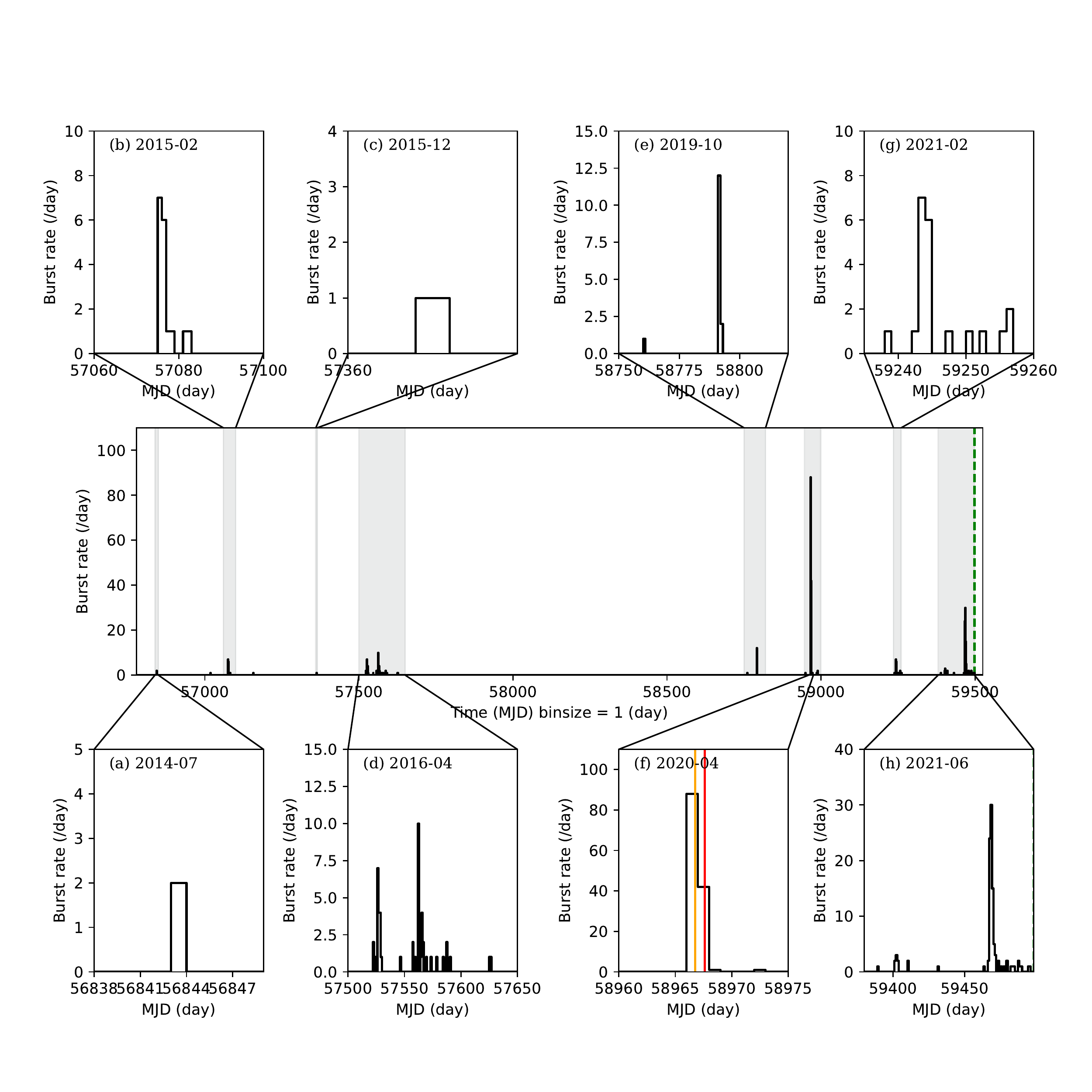}
 \caption{The overall burst rate of our sample from 2014 May 1 to 2021 July 8. The details of the eight bursting phases are shown in the zoomed panels (a, b, c, d, e, f, g, and h). In panel (f), the yellow line represents the burst forest on 2020 April 27. The red line represents the arrival time of SGR-FRB 200428. The {dashed} green line indicates the last time we checked on the data for this work, which is 2021 October 21.}
\end{figure*}

\end{enumerate}

We obtained {356} burst candidates after the above three steps. Those candidates were further screened manually to ensure no previous study had claimed they were from other nearby sources. We found that more than {65\%} of the candidates in our sample have been previously studied or collected in the literature \citep[e.g.,][]{2020ApJ...902L..43L,2020ApJ...893..156L,2020Natur.587...63L} and various resources (such as the IPN SGR Burst list\footnote{\url{ http://www.ssl.berkeley.edu/ipn3/sgrlist.txt}}) as bursts of SGR J1935+2154, which confirms our results. On the other hand, a burst found at 2019-11-14T19:50:42 was excluded according to the IPN SGR Burst list, which considered the burst as one from SGR 1900+14. In addition, two GRBs (GRB 190619B and GRB 201218A) were also excluded after cross-checking the Fermi GBM GRB catalog\footnote{\url{https://heasarc.gsfc.nasa.gov/FTP/fermi/data/gbm/bursts}}. We finally obtained {353} bursts in our sample, which, based on our analysis, are associated with SGR J1935+2154.

All the bursts of our results are listed in Table \ref{tab:catalog}, and highlighted {via their burst rate (with a bin size of one day)} in the timeline in Figure \ref{fig:summary}. We note that there are eight bursting phases, starting from 2014 July, 2015 February, 2016 May, 2019 October, 2020 April, 2021 February, and 2021 June, respectively. 
We also emphasize that there was no burst found in our search in 2017 and 2018, although the sky coverage of Fermi/GBM data was as usual.

\begin{figure}
 \label{fig:long-GRB-like}
 \centering
 \includegraphics[width=0.49\textwidth]{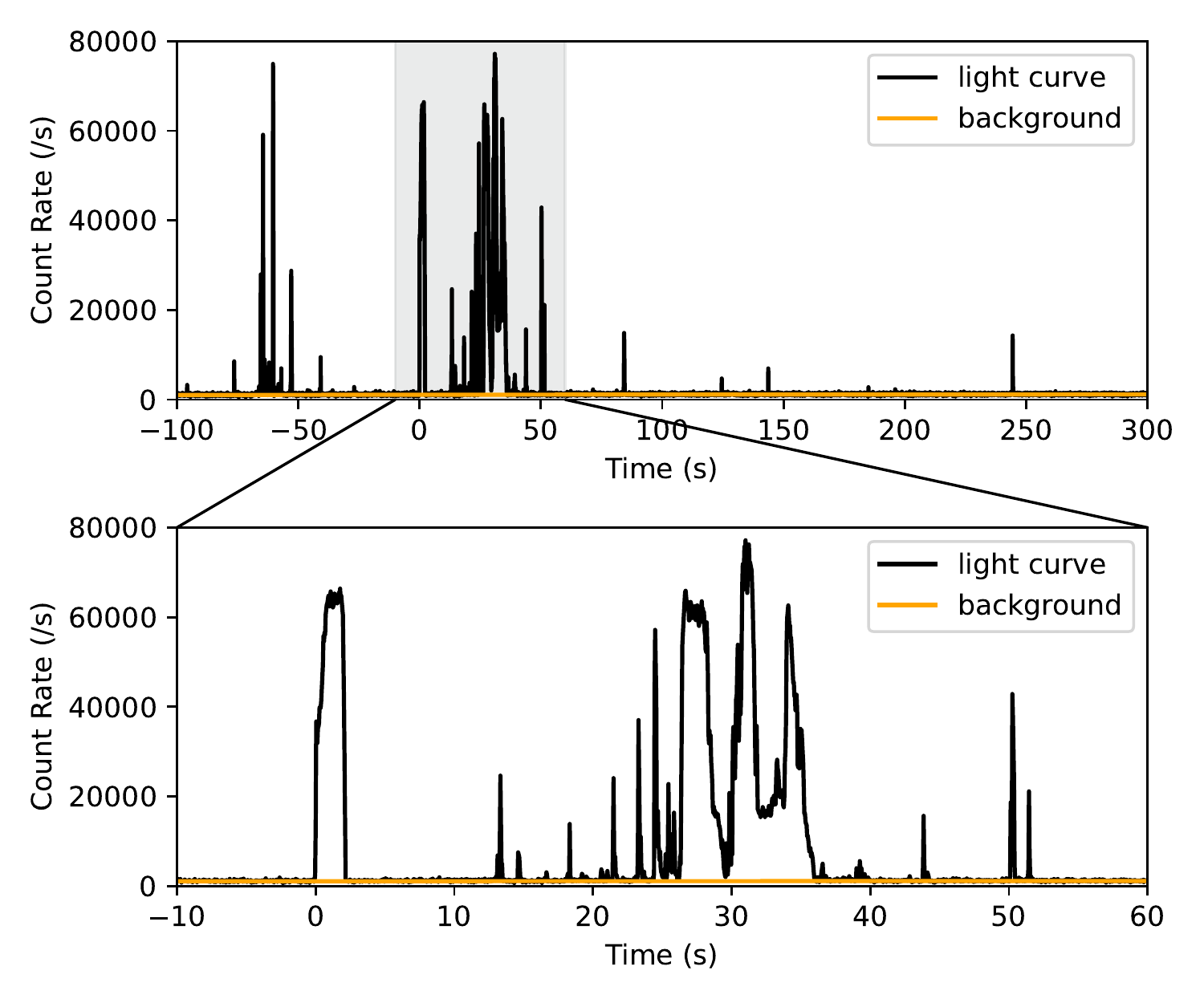}
 \caption{Light curves of the burst forest on 2020 April 27. The yellow curve represents the background. The time corresponding to $T=0$ is 2020-04-27T18:32:41:650 (UTC).}
\end{figure}

Most bursts of our sample are short-duration bursts (see Figure \ref{fig:eg_lc} for an example). A burst forest was observed on 2020 April 27, as shown in Figure \ref{fig:long-GRB-like}. We also identified two intermediate bursts (IBs) at 2020-04-27T18:32:41.640 (UTC) and 2020-04-27T18:33:05.840 (UTC), which are characterized by their relatively long durations compared to short-duration bursts. 
No giant flare was found in our results. About 20 hr later, the IB and burst forest was followed by the X-ray burst associated with a fast radio burst, FRB 200428 (the red vertical line in Figure \ref{fig:summary}).

\section{Periodicity Analysis}\label{sec:physical_origin}

\begin{figure}
 \centering
 \includegraphics[width=0.47\textwidth]{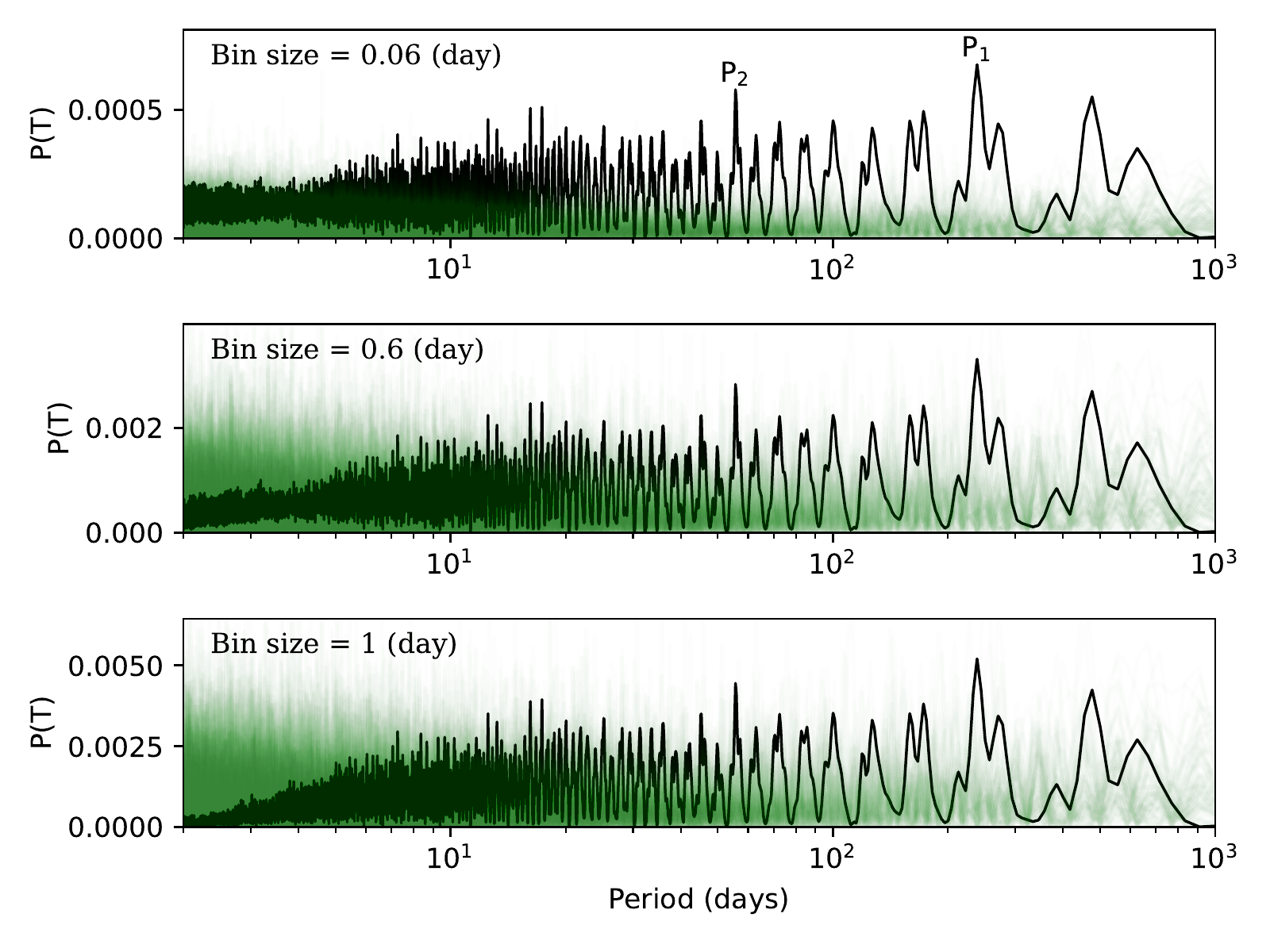} 
 \caption{{The Lomb-Scargle periodograms with different bin sizes. There are two significant peaks at Period $\sim$ 238 days ($P_1$), and Period $\sim$ 55 days ($P_2$). The green lines overplotted in the background represent the periodograms of the 10000 simulated data sets.}}
 \label{fig:lombscargle1}
\end{figure}

The complete seven-phase, 353-burst events obtained from our data reduction above provide a rich and homogeneous sample to perform the PWB analysis of this magnetar. To do so, we employed the Lomb-Scargle (LS) method\footnote{\url{https://docs.astropy.org/en/stable/timeseries/lombscargle.html}}, which is a widely used approach to derive the period of unevenly sampled observations \citep{Lomb1976Ap&SS..39..447L, Scargle1982ApJ...263..835S, VanderPlas2018ApJS..236...16V}. {To utilize the LS method, one has to bin the observed events into uniform bins and obtain the event rate as a function of time. Based on the various event rate binned with values between 0.06 and 1 day, we calculated their Lomb-Scargle periodograms of SGR J1935+2154, as shown in Figure \ref{fig:lombscargle1}.} 
The most significant peak in the periodograms is at \scalgleP\ days (P${_1}$), which is roughly consistent with the previous study by \citet{2021PASP..133g4202G}, which proposed a period of about 231 days using a 161-burst sample observed by the IPN network from 2014 to 2020. {We noticed that there is one less significant peaks presented in Figure \ref{fig:lombscargle1}, namely at $\sim$55 days ($P_2$). Both $P_1$ and $P_2$ are subject to further significance check before they can be claimed as a period of the SGR.}

\begin{figure}
 \centering
 \includegraphics[width=0.47\textwidth]{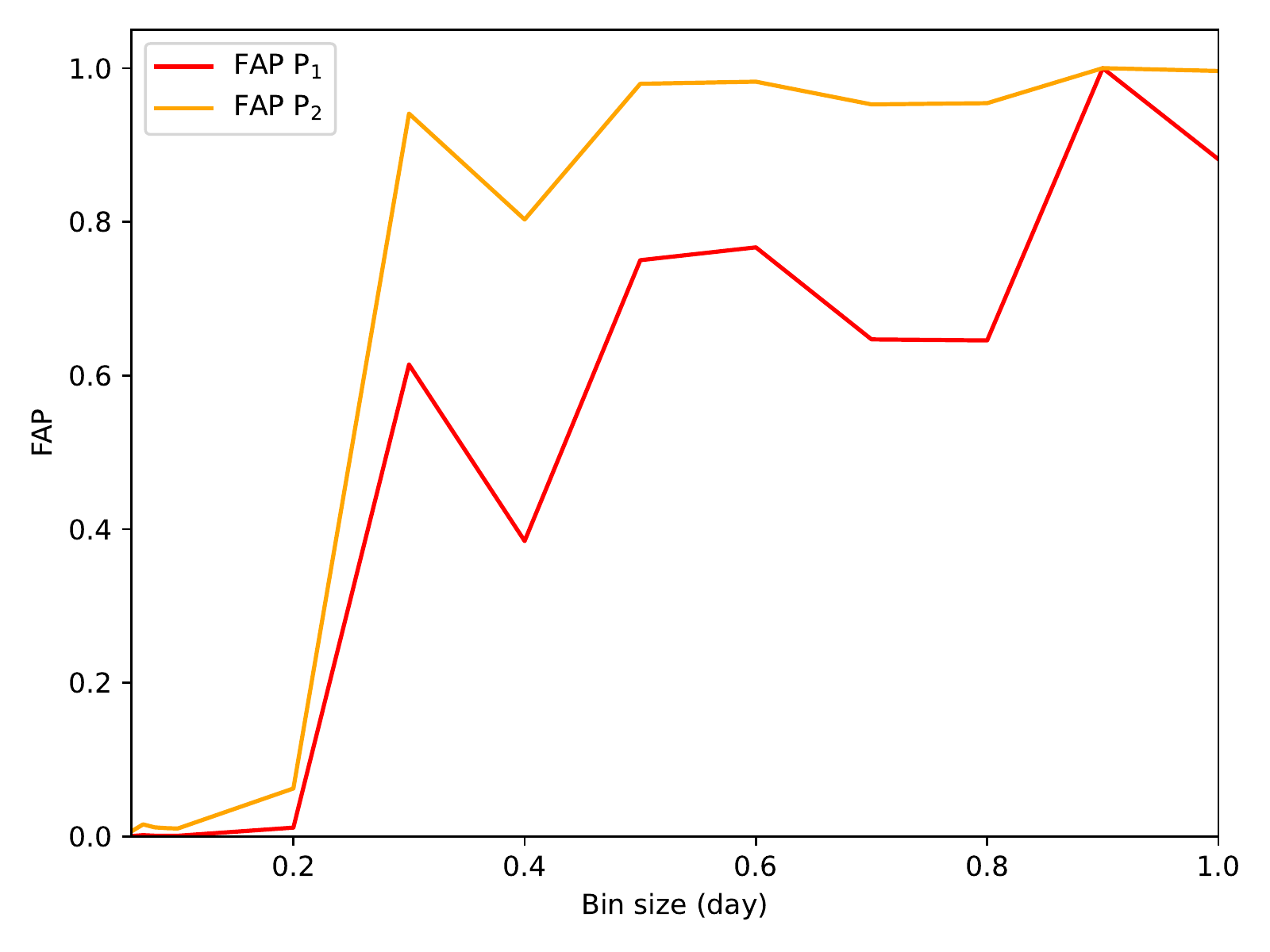} 
 \caption{{The effect of bin size on FAP. The blue and yellow lines represent the FAPs of $P_1$ and $P_2$.}}
 \label{fig:fapbins}
\end{figure}

\begin{figure}
 \centering
 \includegraphics[width=0.47\textwidth]{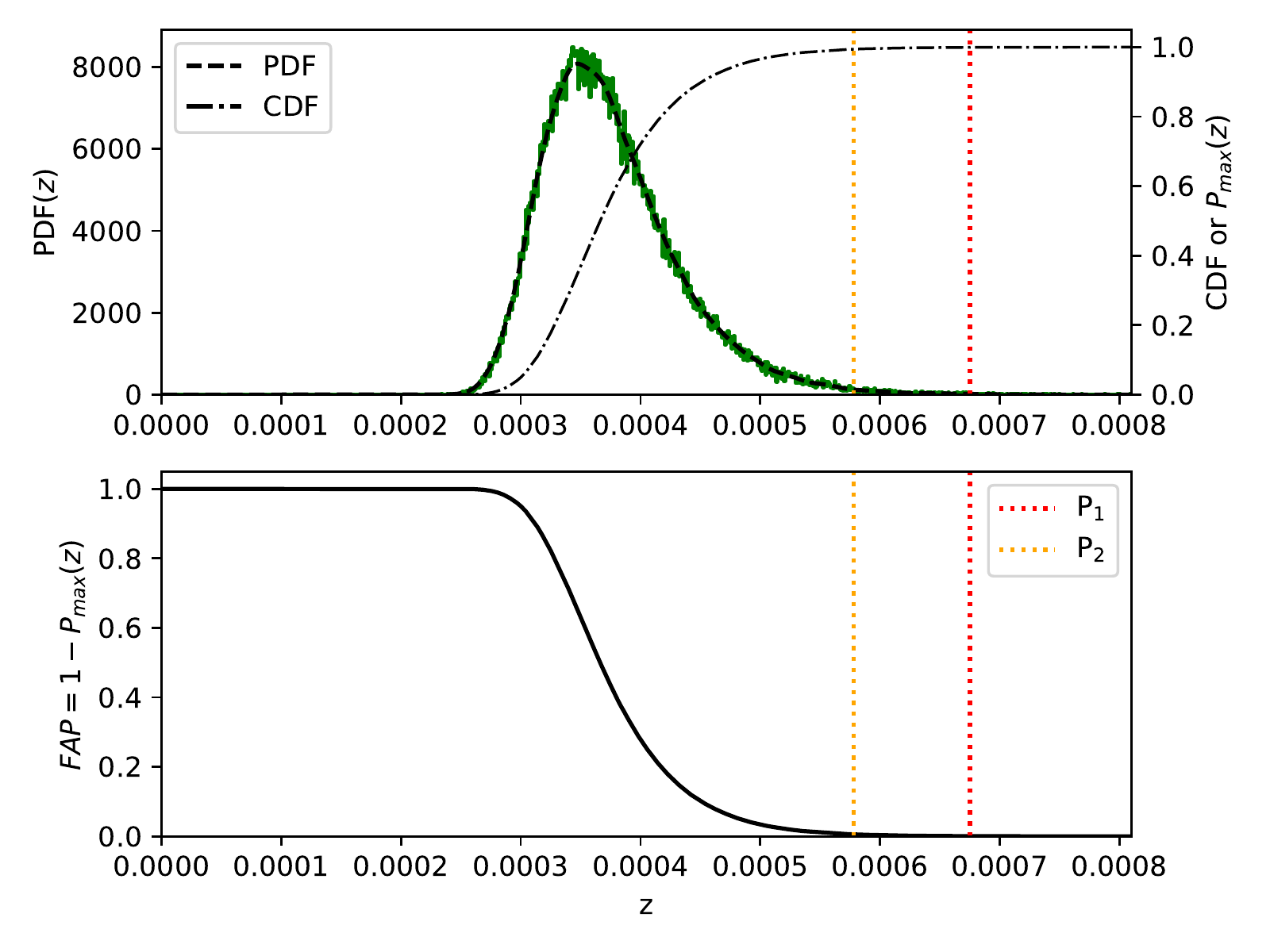} 
 \caption{{The top panel shows the probability density function (PDF) and cumulative distribution function (CDF) of $z$ (taken under the null hypothesis). The bottom panel shows the FAP curve. The red and yellow dotted lines represent $P_1$ and $P_2$. The bin size is 0.06 day.}}
 \label{fig:fap}
\end{figure}

{
The typical way of quantifying the significance of a peak in a periodogram is the false alarm probability \citep[FAP,][]{Scargle1982ApJ...263..835S}, which measures the probability that a data set with no signal would lead to a peak with a similar magnitude \citep{VanderPlas2018ApJS..236...16V}. 
\cite{2008MNRAS.385.1279B} derived an analytic FAP based on the theory of extreme values for stochastic processes in the form of
\begin{equation}
FAP(z) = 1-P_{max}(z,f_{max}),
\label{eq:fap}
\end{equation}
where $z$ is the height of a peak in periodogram, $f_{max}$ is the maximum frequency of the calculated periodogram, and $P_{max}(z,f_{max})$ denotes the cumulative distribution function of the maximum of z under the null hypothesis in the frequency range between 0 and $f_{max}$. FAP calculation depends on the size of the input data set, and in our case, the bin size of sampling to obtain the event rate \citep{VanderPlas2018ApJS..236...16V}. To illustrate this, we calculated the FAP values following Eq. \ref{eq:fap} for the periodograms of the data sets with different bin sizes, as shown in Figure \ref{fig:fapbins} and Table \ref{tab:fap}. We noticed that the FAP values become stable when the bin size $\le $ 0.06 day, with typical values of FAP({P$_{1}$}) = $7.7\times 10^{-4}$ and FAP($P_2$)= $7.0\times 10^{-3}$. Our calculation suggests that a proper sampling bin size (e.g., $\le $ 0.06 day or $\sim$ 5000 s) is crucial to reflect the temporal structure of the events and to claim the significance of the periodic signal.}

{Alternatively, one can calculate the FAP through a Monte Carlo simulation. To do so, we simulated 100,000 sets of events. In each set, there are 353 bursts randomly distributed from 2013 January to 2021 October. We then calculate the LS periodogram for each set. By measuring the numbers of cases, N, out of the 100,000 simulations, in which one can reproduce the same or higher height z of the peaks (i.e., $P_1$ and $P_2$) in the observed data, we can calculate the FAP as FAP$=\frac{N}{100000}$. Our simulations are shown in Figure \ref{fig:fap} and yield a result of FAP({P$_{1}$}) = $6.8\times 10^{-4}$ and FAP($P_2$)$= 5.9\times 10^{-3}$, which is consistent with the result obtained by Eq. \ref{eq:fap}. Based on the above calculation, we claim $P_1$ is a period of SGR 1935+2154 at $4.5\sigma$ confidence level. On the other hand, $P_2$ is much less significant with a confidence level less than $4\sigma$. Furthermore, as we show below (Section \ref{sec:summary}) that $P_2$ becomes unstable when considering the data gaps, and thus can be a false positive.}

\begin{figure}
\centering
 \label{fig:fold}
 \includegraphics[width=0.45\textwidth]{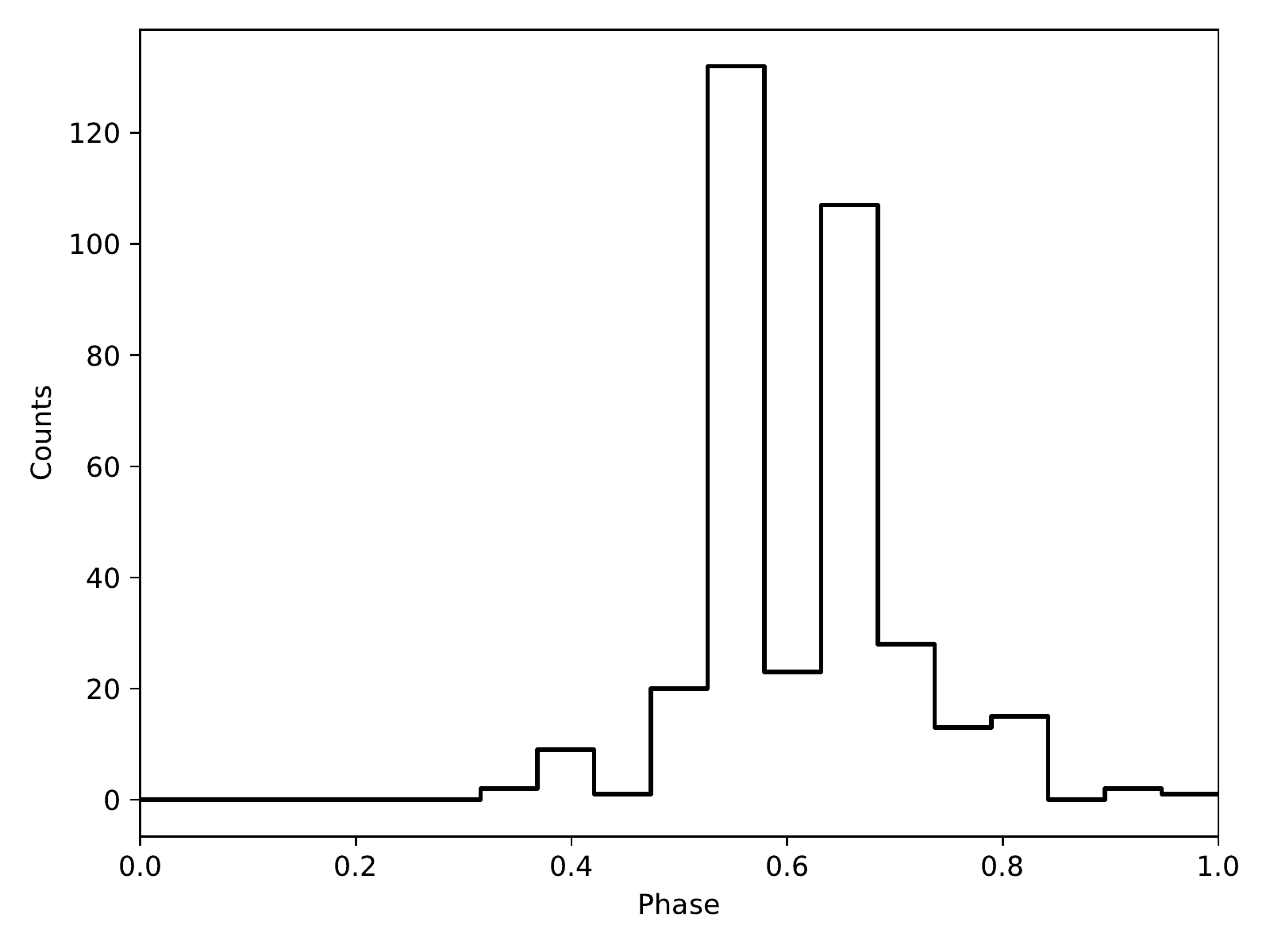} 
 \caption{{Phase-folded burst rate according to the 238 day period. The duty cycle is determined by the width of the  folded data.}}
\end{figure}

\begin{figure}
\centering
 \label{fig:burst_rate}
 \includegraphics[width=0.45\textwidth]{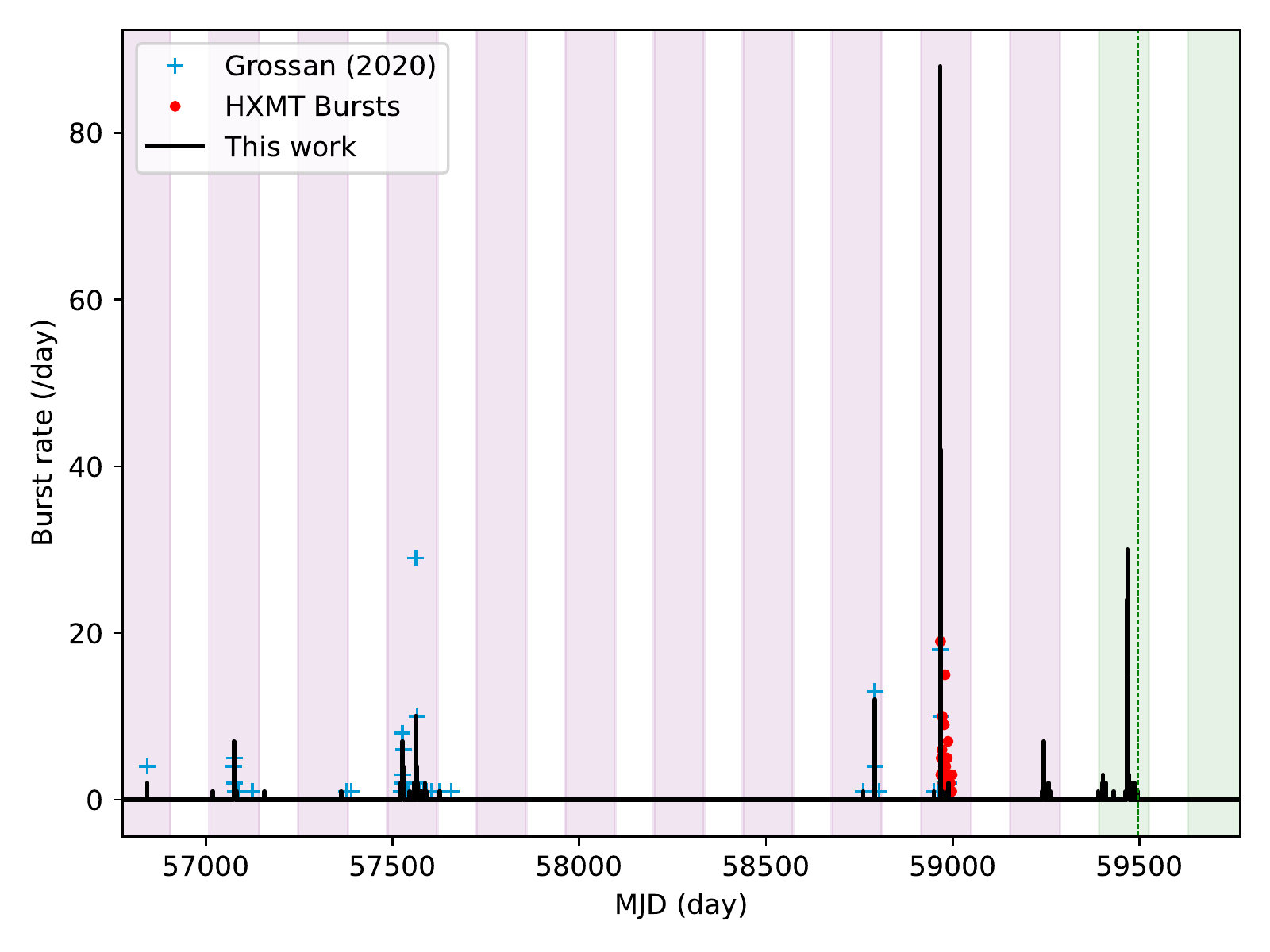}
 \caption{{The period plot according to $P_1$. The shaded areas show the active windows, among which the green ones show the predicted two nearest upcoming active windows.} Three different samples, this work (black lines), \cite{2021PASP..133g4202G} (blue points), and HXMT bursts (red points) are overplotted with different symbols. The dotted green line indicates the current date, which is 2021 October 21. }
\end{figure}

{Using a period of $P_1$,  we plot the phase-folded event rate in Figure \ref{fig:fold}, from which we can further calculate a duty cycle to be 63.2\%.   The active windows from our calculations are marked with pink regions in Figure \ref{fig:burst_rate}, where we also overplot the Fermi bursts in our sample as well as those IPN bursts from \citep{2021PASP..133g4202G} and the bursts from the HXMT mission\footnote{\url{http://hxmtweb.ihep.ac.cn/bursts/392.jhtml}}, with the latter two not being taken into account in our fit. One can see that all observed data fully comply with our model's prediction.  }

 \section{Implications and Discussions} \label{sec:summary}

By analyzing a complete {353}-burst sample searched for periodic signal from the 8.5yr up-to-date continuous event data of Fermi/GBM mission using {the LS} methods, we identified a period of \zjhP\ days with a {63.2\% duty cycle}. Our model suggested a total of 12 cycles from 2014 July to date. For such a long time span, the time clustering behavior of the bursts is statistically significant \citep{2021PhRvD.104b3007D}. Our results are fully consistent with all the X-ray bursts of SGR J1935+2154 observed by multiple missions to date. {Our calculation} can predict the next active windows in the nearest future, as listed in Table \ref{tab:forecast} and overplotted in Figure \ref{fig:burst_rate}. Interestingly, {as of October 10}, the current ongoing burst activities of the SGR J1935+2154, which starts from 2021 June 26, are fully within our model-predicted window.

\begin{figure}
 \centering
 \includegraphics[width=0.47\textwidth]{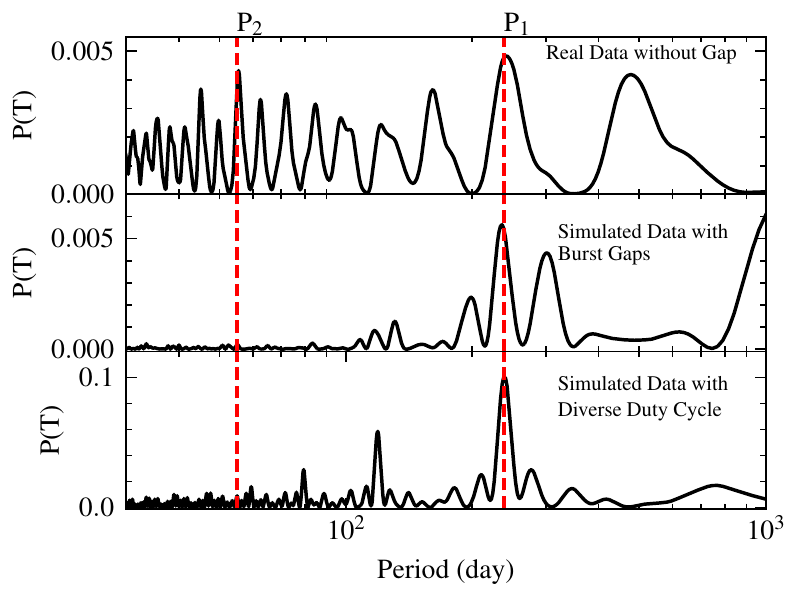}
 \caption{
The Lomb-Scargle periodogram of the real data without the burst gaps is shown in the top panel. 
 The periodograms of the simulated data are shown in the middle and bottom panels. The dashed vertical red lines mark the two periods. The middle panel is the periodogram of the simulated data with the burst gaps. The bottom panel is the periodogram of the simulated data with different active window widths. }
 \label{fig:simulationData3}
\end{figure}

There is a less significant period of $P_2=55 $ day as presented in Figure \ref{fig:lombscargle1}.
We investigate their possible causes through the following analysis:

\begin{enumerate}

 \item We ignore the burst gaps in 2017 and 2018 and recalculate the Lomb-Scargle periodogram from the real data, as presented in the top panel of Figure \ref{fig:simulationData3}. {Although the peak of $P_2$ is visible, it becomes much less significant ($<3\sigma$), suggesting that it can be caused the data gaps.}

 \item Based on the observed data, a series of points that are distributed with a period of \scalgleP \ days and the duty cycle of {63\%} are simulated. The observation-based burst gaps are then applied to the simulated data. The calculated Lomb-Scargle periodogram is shown in the middle panel of Figure \ref{fig:simulationData3}. {The peak of $P_1$ persists in this simulation, suggesting that $P_1$ is not affected by the gap of data.}

 \item Similar to the above step, but data are simulated with different duty cycles but no data gaps. The peak of $P_1$ still persists in this simulation, suggesting that it is not affected by the burst active window itself.

\end{enumerate}

Based on our simulation, the period of {$P_1=238$ days} seems to be always stable and reliable. {On the other hand, the data gaps of the data may lead to some faked signals which should be studied with caution.}

The physical origin of the $P\sim \zjhP$ days, however, remains an open question. Given that there is no evidence showing SGR J1935+2154 are in a binary system, we focus on the explanations invoking the properties of the magnetar itself. In that regard, the most natural way to cause a period may be the free precession of the magnetar. The NS precession period is derived to be $P_{\rm prec} \sim P_{\rm spin} / \epsilon$ \citep{Levin2020ApJ...895L..30L}, where $P_{\rm spin}$ and $\epsilon$ are the spin period and the ellipticity of the NS, respectively. By substituting {$P_{\rm prec}=\zjhP$ days} and $P_{\rm spin} = 3.24 
$ s \citep{2016MNRAS.457.3448I} for SGR J1935+2154, we can derive that $\epsilon \simeq 1.6\times 10^{-7}$. We can further calculate that the ratio of the poloidal component’s energy to the total
magnetic energy is $\Lambda \simeq $ 0.83 \citep[Equation~(7) of ][]{2013MNRAS.434.1658M} for a simple dipole poloidal-plus-toroidal
magnetic field configurations, which suggests that the poloidal component is more dominant than the toroidal component. Interestingly, two FRBs also show periodic behavior. Considering the association between FRB 200428 and the X-ray burst of SGR 1935+2154, their physical origins of periodicity may have a close connection. Therefore, extensive study of the periodicity in X-ray bursts is important to understand the origin of periodic FRBs.

One challenge to our model is that there is no burst observed in 2017 and 2018. Such a gap covers four continuous burst-absent periods. Interestingly, we found that the four periods before and after the gap are both burst-present (Figure \ref{fig:burst_rate}). This suggests that there may be a superposed {$ 8 \times P\sim 1904$ day period}. Such a larger period may be related to the globally evolving collimation of the emission region of SGR J1935+2154. 
We note that this hypothesis can be tested by checking the (non)presence of bursts in our model-predicted window starting in 2022 February (Table \ref{tab:forecast}).

Our model can also back-predict the active windows. The nearest predicted active window before the first discovery of SGR J1935+2154 is between 2013  September and 2014 January, or between 2009 January and 2009 May if considering the superposed $ 8 \times P$ period. Both of them are covered by the IPN network and Fermi. However, neither the IPN list \citep[Confrimed SGR Burst \& Possible SGR Bursts; \url{http://www.ssl.berkeley.edu/ipn3/masterli.html}; ][]{2009AIPC.1133...55H} nor our search has yielded any burst in those windows, even though IPN has operated largely continuously since the year 1990. The nondetection of SGR J1935+2154 in pre-discovery data suggests that SGR J1935+2154 likely began its active phase around 2014 July.

\acknowledgments

 B.B.Z acknowledges
 support by Fundamental Research Funds for the Central Universities
 (14380046), the National Key Research and Development Programs of China
 (2018YFA0404204), the National Natural Science Foundation of China (Grant
 Nos. 11833003, U2038105), the science research grants from the China Manned Space Project with NO.CMS-CSST-2021-B11, and the Program for Innovative Talents,
 Entrepreneur in Jiangsu. F.Y.W. acknowledges support by the National Natural Science
Foundation of China (grant U1831207). We acknowledge the use of public data from the
Fermi Science Support Center (FSSC).

\textit{Software:} {
numpy; matplotlib; pandas; scipy; astropy;
baseline\footnote{\url{https://www.rdocumentation.org/packages/baseline/versions/1.3-1}}; 
sigma\_clip\footnote{\url{https://docs.astropy.org/en/stable/api/astropy.stats.sigma\_clip.html}};
bayesian\_blocks\footnote{\url{https://docs.astropy.org/en/stable/api/astropy.stats.bayesian\_blocks.html\#astropy.stats.bayesian_blocks}};
gbm\_drm\_gen \footnote{\url{https://github.com/grburgess/gbm_drm_gen}}
cartopy\footnote{{\url{https://scitools.org.uk/cartopy/docs/latest/}}};
lomb-scargle\footnote{\url{https://docs.astropy.org/en/stable/timeseries/lombscargle.html}};
pymultinest\footnote{\url{https://github.com/JohannesBuchner/PyMultiNest}}; 
}

\clearpage
\begin{longtable*}{cc|cc|cc}
\caption{The sample of the X-Ray bursts of SGR J1935+2154 }
\label{tab:catalog}\\
\hline 
\hline 
Burst Time (UTC)	&	Duration (s)	& Burst Time (UTC)&	Duration (s)	&	Burst Time (UTC)	&	Duration (s)	\\
 & 8-900 keV & & 8-900 keV & & 8-900 keV \\
\hline 
\endfirsthead

\hline 
\hline 
Burst Time (UTC)	&	Duration (s)	& Burst Time (UTC)&	Duration (s)	&	Burst Time (UTC)	&	Duration (s)	\\
 & 8-900 keV & & 8-900 keV & & 8-900 keV \\
\hline 
\endhead
\bottomrule
\hline 
\endfoot
2014-07-05T09:32:48.640 & 0.12 & 2016-06-23T15:16:09.040 & 0.12 & 2020-04-27T18:31:39.240 & 1.32\\
2014-07-05T09:37:34.480 & 0.04 & 2016-06-23T15:16:26.840 & 0.20 & 2020-04-27T18:31:41.080 & 0.44\\
2014-12-27T02:16:21.280 & 0.12 & 2016-06-23T16:17:04.200 & 0.08 & 2020-04-27T18:31:42.120 & 0.24\\
2015-02-22T13:33:00.000 & 0.12 & 2016-06-23T16:49:57.480 & 0.12 & 2020-04-27T18:31:44.640 & 0.16\\
2015-02-22T14:33:20.240 & 0.28 & 2016-06-23T17:39:22.360 & 0.08 & 2020-04-27T18:31:45.520 & 0.08\\
2015-02-22T17:57:05.960 & 0.08 & 2016-06-23T17:55:48.800 & 0.08 & 2020-04-27T18:31:48.360 & 0.76\\
2015-02-22T19:24:45.960 & 0.04 & 2016-06-23T19:24:40.040 & 0.08 & 2020-04-27T18:32:00.520 & 0.52\\
2015-02-22T19:44:16.880 & 0.08 & 2016-06-23T19:36:27.360 & 0.20 & 2020-04-27T18:32:05.840 & 0.04\\
2015-02-22T19:58:16.640 & 0.16 & 2016-06-23T19:38:00.000 & 0.12 & 2020-04-27T18:32:14.680 & 0.44\\
2015-02-22T23:57:33.880 & 0.12 & 2016-06-23T20:06:37.120 & 0.16 & 2020-04-27T18:32:30.680 & 0.64\\
2015-02-23T01:38:07.960 & 0.08 & 2016-06-25T08:04:52.960 & 0.08 & 2020-04-27T18:32:31.520 & 0.68\\
2015-02-23T05:05:06.520 & 0.08 & 2016-06-26T06:03:14.960 & 0.32 & 2020-04-27T18:32:39.000 & 0.08\\
2015-02-23T05:24:53.840 & 0.52 & 2016-06-26T13:54:30.720 & 0.84 & 2020-04-27T18:32:41.640 & 2.20\\
2015-02-23T06:45:40.080 & 0.08 & 2016-06-26T17:50:03.240 & 0.04 & 2020-04-27T18:32:54.480 & 0.68\\
2015-02-23T14:26:16.240 & 0.08 & 2016-06-26T20:34:49.320 & 0.08 & 2020-04-27T18:32:56.240 & 0.20\\
2015-02-23T16:13:52.480 & 0.04 & 2016-06-27T01:50:15.840 & 0.08 & 2020-04-27T18:32:58.320 & 0.04\\
2015-02-24T06:48:47.160 & 0.08 & 2016-06-27T09:44:07.720 & 0.04 & 2020-04-27T18:32:59.680 & 0.36\\
2015-02-25T10:51:03.600 & 0.12 & 2016-06-30T10:02:26.000 & 0.08 & 2020-04-27T18:33:00.840 & 0.48\\
2015-02-28T10:56:48.800 & 0.08 & 2016-07-04T14:33:38.640 & 0.12 & 2020-04-27T18:33:02.240 & 0.16\\
2015-03-01T07:28:54.120 & 0.08 & 2016-07-09T16:54:28.200 & 0.32 & 2020-04-27T18:33:02.720 & 0.68\\
2015-05-15T18:16:51.280 & 0.16 & 2016-07-15T07:09:11.720 & 0.08 & 2020-04-27T18:33:04.560 & 0.04\\
2015-12-06T23:49:57.240 & 0.04 & 2016-07-18T09:36:06.520 & 0.12 & 2020-04-27T18:33:04.640 & 1.08\\
2016-05-14T08:21:54.560 & 0.16 & 2016-07-18T09:36:07.120 & 0.08 & 2020-04-27T18:33:05.840 & 12.72\\
2016-05-14T22:25:21.840 & 0.04 & 2016-07-21T09:36:13.680 & 0.20 & 2020-04-27T18:33:24.320 & 0.24\\
2016-05-16T20:49:47.000 & 0.04 & 2016-08-25T07:26:53.480 & 0.44 & 2020-04-27T18:33:25.480 & 0.44\\
2016-05-18T07:49:34.000 & 0.20 & 2016-08-26T14:05:52.920 & 0.16 & 2020-04-27T18:33:31.760 & 0.32\\
2016-05-18T09:09:23.880 & 0.20 & 2019-10-04T09:00:53.600 & 0.12 & 2020-04-27T18:33:33.000 & 0.52\\
2016-05-18T10:07:26.760 & 0.04 & 2019-11-04T01:20:24.000 & 0.20 & 2020-04-27T18:33:53.120 & 0.24\\
2016-05-18T10:28:02.760 & 0.16 & 2019-11-04T02:53:31.360 & 0.08 & 2020-04-27T18:34:05.720 & 0.44\\
2016-05-18T15:33:47.000 & 0.08 & 2019-11-04T04:26:55.880 & 0.08 & 2020-04-27T18:34:46.040 & 0.24\\
2016-05-18T17:00:31.720 & 0.04 & 2019-11-04T07:20:33.720 & 0.12 & 2020-04-27T18:34:47.240 & 0.32\\
2016-05-18T19:40:37.080 & 0.60 & 2019-11-04T09:17:53.480 & 0.64 & 2020-04-27T18:34:57.400 & 0.44\\
2016-05-19T11:46:52.480 & 0.12 & 2019-11-04T10:44:26.280 & 0.20 & 2020-04-27T18:35:05.320 & 0.20\\
2016-05-19T11:59:32.600 & 0.08 & 2019-11-04T12:38:38.520 & 0.16 & 2020-04-27T18:35:46.640 & 0.08\\
2016-05-19T12:07:46.600 & 0.12 & 2019-11-04T15:36:47.440 & 0.40 & 2020-04-27T18:35:57.640 & 0.04\\
2016-05-19T19:59:54.960 & 0.32 & 2019-11-04T20:13:42.560 & 0.04 & 2020-04-27T18:36:45.400 & 1.00\\
2016-05-20T03:24:12.840 & 0.04 & 2019-11-04T20:29:39.760 & 0.32 & 2020-04-27T18:38:20.200 & 0.16\\
2016-05-20T05:21:33.480 & 0.16 & 2019-11-04T23:16:49.560 & 0.08 & 2020-04-27T18:38:53.720 & 0.24\\
2016-05-20T16:21:43.160 & 0.12 & 2019-11-04T23:48:01.360 & 0.24 & 2020-04-27T18:40:15.040 & 0.48\\
2016-05-20T21:42:29.320 & 0.28 & 2019-11-05T06:11:08.600 & 0.32 & 2020-04-27T18:40:32.280 & 0.40\\
2016-05-21T03:23:36.640 & 0.12 & 2019-11-05T07:17:17.880 & 0.08 & 2020-04-27T18:40:33.120 & 0.40\\
2016-06-07T03:48:59.400 & 0.04 & 2020-04-10T09:43:54.280 & 0.20 & 2020-04-27T18:42:40.800 & 0.08\\
2016-06-18T01:42:55.560 & 0.08 & 2020-04-27T18:26:20.160 & 0.12 & 2020-04-27T18:42:50.680 & 0.32\\
2016-06-18T20:27:25.760 & 0.12 & 2020-04-27T18:31:05.800 & 0.20 & 2020-04-27T18:44:04.640 & 0.72\\
2016-06-20T15:16:34.840 & 0.28 & 2020-04-27T18:31:25.240 & 0.20 & 2020-04-27T18:46:08.760 & 0.24\\
2016-06-22T13:45:23.680 & 0.16 & 2020-04-27T18:31:33.800 & 5.00 & 2020-04-27T18:46:39.440 & 0.04\\
2020-04-27T18:46:40.160 & 0.28 & 2020-04-28T00:46:00.040 & 0.80 & 2021-01-30T03:24:38.360 & 0.12\\
2020-04-27T18:46:40.760 & 0.32 & 2020-04-28T00:46:06.440 & 0.04 & 2021-01-30T08:39:53.840 & 0.16\\
2020-04-27T18:47:05.760 & 0.12 & 2020-04-28T00:46:17.960 & 0.32 & 2021-01-30T10:35:35.120 & 0.08\\
2020-04-27T18:48:38.680 & 0.08 & 2020-04-28T00:46:20.200 & 0.68 & 2021-01-30T17:40:54.760 & 0.16\\
2020-04-27T18:49:28.040 & 0.64 & 2020-04-28T00:46:23.560 & 0.84 & 2021-01-30T21:01:22.840 & 0.12\\
2020-04-27T19:36:05.120 & 0.04 & 2020-04-28T00:46:43.080 & 0.52 & 2021-02-02T12:54:26.960 & 0.24\\
2020-04-27T19:37:39.320 & 0.76 & 2020-04-28T00:47:57.560 & 0.20 & 2021-02-05T07:13:15.760 & 0.08\\
2020-04-27T19:43:44.560 & 0.52 & 2020-04-28T00:48:49.280 & 0.60 & 2021-02-07T00:35:52.720 & 0.12\\
2020-04-27T19:45:00.480 & 0.12 & 2020-04-28T00:49:00.320 & 0.12 & 2021-02-10T14:15:07.080 & 0.08\\
2020-04-27T19:55:32.320 & 0.56 & 2020-04-28T00:49:01.120 & 0.24 & 2021-02-11T12:28:00.120 & 0.04\\
2020-04-27T20:01:45.800 & 0.40 & 2020-04-28T00:49:01.960 & 0.16 & 2021-02-11T13:43:16.760 & 0.08\\
2020-04-27T20:13:38.280 & 0.08 & 2020-04-28T00:49:06.480 & 0.08 & 2021-02-16T22:20:39.600 & 0.36\\
2020-04-27T20:14:51.440 & 0.04 & 2020-04-28T00:49:16.640 & 0.28 & 2021-06-24T02:34:10.200 & 0.08\\
2020-04-27T20:15:20.640 & 1.28 & 2020-04-28T00:49:22.400 & 0.12 & 2021-07-06T03:50:09.720 & 0.24\\
2020-04-27T20:17:09.160 & 0.16 & 2020-04-28T00:49:27.320 & 0.08 & 2021-07-06T15:01:52.600 & 0.24\\
2020-04-27T20:17:27.320 & 0.12 & 2020-04-28T00:49:46.160 & 0.04 & 2021-07-07T00:33:31.640 & 0.16\\
2020-04-27T20:17:50.440 & 0.28 & 2020-04-28T00:49:46.680 & 0.20 & 2021-07-07T15:45:11.080 & 0.20\\
2020-04-27T20:17:51.360 & 0.44 & 2020-04-28T00:50:01.040 & 0.52 & 2021-07-07T18:08:16.920 & 0.08\\
2020-04-27T20:17:58.440 & 0.16 & 2020-04-28T00:50:22.000 & 0.04 & 2021-07-08T00:18:18.560 & 0.32\\
2020-04-27T20:19:48.400 & 0.08 & 2020-04-28T00:51:35.920 & 0.08 & 2021-07-08T09:03:41.640 & 0.08\\
2020-04-27T20:19:49.480 & 0.20 & 2020-04-28T00:51:55.440 & 0.12 & 2021-07-15T06:43:16.360 & 0.12\\
2020-04-27T20:21:51.840 & 0.04 & 2020-04-28T00:52:06.240 & 0.04 & 2021-07-15T09:15:18.840 & 0.12\\
2020-04-27T20:21:52.280 & 0.16 & 2020-04-28T00:54:57.480 & 0.16 & 2021-08-05T00:08:56.000 & 0.20\\
2020-04-27T20:21:55.160 & 0.60 & 2020-04-28T00:56:49.640 & 0.32 & 2021-09-06T01:44:30.080 & 0.16\\
2020-04-27T20:25:53.480 & 0.36 & 2020-04-28T01:04:03.160 & 0.04 & 2021-09-09T18:57:14.840 & 0.12\\
2020-04-27T21:14:45.600 & 0.32 & 2020-04-28T02:00:11.440 & 0.40 & 2021-09-09T20:21:28.360 & 0.08\\
2020-04-27T21:15:36.400 & 0.16 & 2020-04-28T02:27:24.920 & 0.04 & 2021-09-10T00:45:46.880 & 1.44\\
2020-04-27T21:20:55.560 & 0.12 & 2020-04-28T03:32:00.600 & 0.08 & 2021-09-10T00:46:21.000 & 0.08\\
2020-04-27T21:43:06.320 & 0.24 & 2020-04-28T03:47:52.160 & 0.24 & 2021-09-10T01:00:43.680 & 0.80\\
2020-04-27T21:48:44.080 & 0.44 & 2020-04-28T04:09:47.320 & 0.08 & 2021-09-10T01:04:33.360 & 0.48\\
2020-04-27T21:59:22.520 & 0.24 & 2020-04-28T05:56:30.560 & 0.08 & 2021-09-10T01:06:23.720 & 0.64\\
2020-04-27T22:47:05.360 & 0.52 & 2020-04-28T09:51:04.880 & 0.28 & 2021-09-10T01:08:40.680 & 0.56\\
2020-04-27T22:55:19.920 & 0.28 & 2020-04-29T20:47:28.000 & 0.20 & 2021-09-10T01:13:17.400 & 0.04\\
2020-04-27T23:02:53.520 & 0.60 & 2020-05-03T23:25:13.440 & 0.24 & 2021-09-10T01:13:57.800 & 0.04\\
2020-04-27T23:06:06.160 & 0.16 & 2020-05-16T18:12:52.120 & 0.08 & 2021-09-10T01:14:36.880 & 0.32\\
2020-04-27T23:25:04.360 & 0.52 & 2020-05-19T18:32:30.320 & 0.68 & 2021-09-10T01:17:19.080 & 0.16\\
2020-04-27T23:27:46.320 & 0.08 & 2020-05-20T14:10:49.840 & 0.12 & 2021-09-10T01:18:54.160 & 0.60\\
2020-04-27T23:44:31.800 & 0.72 & 2020-05-20T21:47:07.520 & 0.48 & 2021-09-10T01:20:48.320 & 0.08\\
2020-04-28T00:19:44.200 & 0.08 & 2021-01-24T20:48:23.960 & 0.08 & 2021-09-10T01:21:49.600 & 0.08\\
2020-04-28T00:23:04.760 & 0.16 & 2021-01-28T23:35:02.480 & 0.16 & 2021-09-10T01:31:40.040 & 0.48\\
2020-04-28T00:24:30.320 & 0.28 & 2021-01-29T02:46:22.560 & 0.20 & 2021-09-10T01:34:18.880 & 0.16\\
2020-04-28T00:37:36.160 & 0.12 & 2021-01-29T07:00:01.000 & 0.24 & 2021-09-10T02:21:06.400 & 0.16\\
2020-04-28T00:39:39.600 & 0.56 & 2021-01-29T10:35:39.960 & 0.32 & 2021-09-10T02:36:38.240 & 0.56\\
2020-04-28T00:41:32.160 & 0.44 & 2021-01-29T15:23:29.920 & 0.08 & 2021-09-10T02:44:34.160 & 0.08\\
2020-04-28T00:43:25.200 & 0.48 & 2021-01-29T15:29:23.920 & 0.48 & 2021-09-10T05:35:55.480 & 0.12\\
2020-04-28T00:44:08.240 & 0.40 & 2021-01-29T18:38:04.840 & 0.12 & 2021-09-10T05:40:48.640 & 0.08\\
2020-04-28T00:44:09.280 & 0.28 & 2021-01-29T21:15:55.960 & 0.08 & 2021-09-10T09:12:48.880 & 0.08\\
2020-04-28T00:45:31.160 & 0.08 & 2021-01-30T00:41:51.160 & 0.16 & 2021-09-10T13:40:20.360 & 0.24\\
2021-09-10T15:50:56.880 & 0.08 & 2021-09-11T16:50:03.840 & 0.04 & 2021-09-13T03:29:21.440 & 0.08\\
2021-09-10T23:40:34.440 & 0.32 & 2021-09-11T17:01:09.760 & 1.44 & 2021-09-13T04:52:06.520 & 0.12\\
2021-09-11T02:28:08.400 & 0.08 & 2021-09-11T17:10:48.600 & 0.48 & 2021-09-13T08:16:46.760 & 0.12\\
2021-09-11T02:28:11.680 & 0.08 & 2021-09-11T18:54:36.040 & 0.04 & 2021-09-13T19:51:33.160 & 0.20\\
2021-09-11T03:02:28.320 & 0.08 & 2021-09-11T20:05:46.200 & 0.08 & 2021-09-14T06:12:45.200 & 0.04\\
2021-09-11T05:32:38.640 & 0.28 & 2021-09-11T20:13:40.480 & 0.32 & 2021-09-14T11:10:36.200 & 0.12\\
2021-09-11T10:42:51.840 & 0.08 & 2021-09-11T20:22:58.800 & 1.28 & 2021-09-14T14:44:21.240 & 0.08\\
2021-09-11T11:45:04.960 & 0.04 & 2021-09-11T22:51:41.560 & 0.12 & 2021-09-16T17:28:31.040 & 0.08\\
2021-09-11T11:53:57.280 & 0.08 & 2021-09-12T00:34:37.320 & 0.76 & 2021-09-16T20:24:09.120 & 0.12\\
2021-09-11T13:27:33.840 & 0.08 & 2021-09-12T00:45:49.400 & 0.04 & 2021-09-18T10:39:07.040 & 0.08\\
2021-09-11T13:28:54.960 & 0.20 & 2021-09-12T05:24:05.640 & 0.08 & 2021-09-20T05:39:51.480 & 0.08\\
2021-09-11T13:29:08.200 & 0.04 & 2021-09-12T07:04:53.840 & 0.08 & 2021-09-22T01:40:44.280 & 0.08\\
2021-09-11T14:58:38.600 & 0.04 & 2021-09-12T07:28:07.240 & 0.36 & 2021-09-22T09:57:03.320 & 0.04\\
2021-09-11T15:03:00.400 & 0.20 & 2021-09-12T10:10:11.680 & 0.12 & 2021-09-25T03:03:28.320 & 0.12\\
2021-09-11T15:06:43.200 & 0.44 & 2021-09-12T12:19:20.440 & 0.48 & 2021-09-26T15:08:11.880 & 0.08\\
2021-09-11T15:15:25.400 & 1.20 & 2021-09-12T13:55:16.440 & 0.08 & 2021-09-27T19:46:09.760 & 0.08\\
2021-09-11T15:16:25.720 & 0.08 & 2021-09-12T15:03:50.600 & 0.40 & 2021-09-30T01:31:06.160 & 0.12\\
2021-09-11T15:17:45.280 & 0.92 & 2021-09-12T16:26:08.040 & 0.08 & 2021-09-30T20:41:15.800 & 0.20\\
2021-09-11T15:26:11.560 & 0.12 & 2021-09-12T20:16:10.400 & 1.00 & 2021-10-01T00:04:04.320 & 0.12\\
2021-09-11T15:32:33.400 & 0.36 & 2021-09-12T20:16:20.280 & 0.12 & 2021-10-02T06:08:13.080 & 0.08\\
2021-09-11T15:34:39.960 & 0.28 & 2021-09-12T20:57:28.720 & 0.16 & 2021-10-07T17:39:49.320 & 0.20\\
2021-09-11T15:38:25.680 & 0.16 & 2021-09-12T20:57:29.320 & 0.08 & 2021-10-08T15:57:46.400 & 0.40\\
2021-09-11T16:36:57.840 & 0.12 & 2021-09-12T23:19:32.080 & 0.28 & 2021-10-11T11:59:40.800 & 0.16\\
2021-09-11T16:39:20.920 & 0.08 & 2021-09-13T00:27:24.920 & 0.56 & & \\

\end{longtable*}

\begin{table}[htbp]
\begin{center}
\caption{The FAPs with different bin sizes.}
\label{tab:fap}
\begin{tabular}{l|ll}
\hline
\hline
Bin Size (day)& $P_1$ FAP & $P_2$ FAP \\

\hline

0.06 & 0.00068 & 0.00591 \\
0.07 & 0.00194 & 0.01406 \\
0.08 & 0.00140 & 0.01032 \\
0.09 & 0.00129 & 0.00982 \\
0.10 & 0.00124 & 0.00938 \\
0.20 & 0.01015 & 0.06160 \\
0.30 & 0.61819 & 0.94697 \\
0.40 & 0.38717 & 0.80979 \\
0.50 & 0.75718 & 0.98207 \\
0.60 & 0.77241 & 0.98597 \\
0.70 & 0.64913 & 0.95684 \\
0.80 & 0.64938 & 0.95797 \\
0.90 & 1.00000 & 1.00000 \\
1.00 & 0.89046 & 0.99775 \\

\hline
\hline
\end{tabular}
\end{center}
\end{table}

\begin{table}[htbp]
\begin{center}
\caption{The nearest upcoming active windows predicted by our model.
}
\label{tab:forecast}
\setlength{\tabcolsep}{7mm}{
\begin{tabular}{ll}

\hline
\hline
Starting Time (UTC) & Ending Time (UTC) \\
\hline
2021-06-25 & 2021-11-08 \\
2022-02-19&	2022-07-04 \\
\hline
\hline
\end{tabular}}
\end{center}
\end{table}

\end{document}